\documentclass[10pt,conference]{IEEEtran}
\usepackage{amsmath}
\usepackage{amsthm}
\usepackage{cite}
\usepackage{amssymb}
\usepackage{graphicx}
\usepackage{algorithm}
\usepackage{algorithmic}
\usepackage{balance}

\ifCLASSINFOpdf
\else
\fi

\begin{document}
\title{Data-Driven Modulation Optimization with LMMSE Equalization for Reliability Enhancement in Underwater Acoustic Communications}
\author{
	Xuehan~Wang\textsuperscript{1},~Hengyu~Zhang\textsuperscript{1},~Jintao~Wang\textsuperscript{1,2},~Zhi~Sun\textsuperscript{1},~Bo~Ai\textsuperscript{3}\\
	\IEEEauthorblockA{
		\textsuperscript{1}Beijing National Research Center for Information Science and Technology (BNRist),\\
		Dept. of Electronic Engineering, Tsinghua University, Beijing 100084, China\\
		\textsuperscript{2}State Key laboratory of Space Network
		and Communications, Tsinghua University, Beijing 100084, China\\
		\textsuperscript{3}School of Electronic and Information Engineering, Beijing Jiaotong University, Beijing 100044, China\\
		Email: \{wang-xh21@mails., zhanghen23@mails., wangjintao\}@tsinghua.edu.cn, \\ zhisun@ieee.org, boai@bjtu.edu.cn.}
}
\maketitle
\begin{abstract} 
Ultra-reliable underwater acoustic (UWA) communications serve as one of the key enabling technologies for future space-air-ground-underwater integrated networks. However, the reliability of current UWA transmission is still insufficient since severe performance degradation occurs for conventional multicarrier systems in UWA channels with severe delay-scale spread. To solve this problem, we exploit learning-inspired approaches to optimize the modulation scheme under the assumption of linear minimum mean square error (LMMSE) equalization, where the discrete representation of waveforms is adopted by utilizing Nyquist filters. The optimization problem is first transferred into maximizing the fairness of estimation mean square error (MSE) for each data symbol since the total MSE is invariant considering the property of orthogonal modulation. The Siamese architecture is then adopted to obtain consistent optimization results across various channel conditions, which avoids the overhead of online feedback, cooperation, and deployment of neural networks and guarantees generalization. The overall scheme including the loss function, neural network structure, and training process is also investigated in depth in this paper. The excellent performance\footnote{Open-source codes of this paper are available at https://github.com/neatlyw/UWA\_moduOptimization\_ICCC2025.} and robustness of the proposed modulation scheme are verified by carrying out the bit error rate test over various UWA channels with severe delay-scale spread.
\end{abstract}

\IEEEpeerreviewmaketitle
\section{Introduction}
High-speed and ultra-reliable underwater acoustic (UWA) communications are required to achieve the target of space-air-ground-underwater integrated networks, which is known as one of the key visions of future wireless connections \cite{integrate_UWA_background,shi_future}. However, though significant progress has been achieved for the cellular networks on the ground, UWA links are still under-developed with low data rates and terrible reliability, which require the innovation of the transmission schemes \cite{OHFDM_us}. \par
To enhance the capacity of UWA communications, the multicarrier scheme based on the orthogonal frequency division multiplexing (OFDM) modulation with zero-padding (ZP) was first investigated in \cite{OFDM_UWA_first}. By appropriate resampling and compensation, conventional single-tap frequency domain equalization can be deployed with excellent performance for path-invariant Doppler scaling factors. However, as indicated in \cite{UWA_simulator,ref_TSP_CE_classic,ref_VBMC,ICI_banded}, the realistic UWA channels usually have path-specific delay and Doppler scaling factors, known as the delay-scale spread property. Since prior theoretical analysis has proved that interference-free modulation schemes do not exist in such scenarios \cite{ref_ODSSM,inter_must}, various transmission schemes have been proposed to carry out the interference mitigation \cite{ref_VBMC} or cancellation \cite{ICI_banded}.\par 
Since interference-free transmission cannot be acquired in typical UWA scenarios, interference reconstruction can be helpful in further promoting the performance limitation under known equalizer frameworks. To be more specific, by appropriately reshaping the structure of the equivalent channel, the receiver can recover the information utilizing the received symbols with inter-symbol-interference (ISI). Inspired by this ingenious innovation, plenty of waveform schemes have been developed for radio-frequency (RF) scenarios, such as the orthogonal time frequency space (OTFS) \cite{ref_OTFS} and orthogonal delay-Doppler division multiplexing (ODDM) \cite{ref_ODDM,ODDM_detector_kehan} modulation. However, both schemes regarded the Doppler effects as the frequency shift rather than time domain contractions or dilations, which are not suitable for UWA communications \cite{OTFS_DSE_TWC,UWA_OTFS_mine}. The authors in \cite{ref_ODSSM} considered the waveform design to realize appropriate interference reconstruction for UWA channels with time scaling effects. However, the lack of consideration of suitable passband processing and implementation issues made it hard to deploy in realistic UWA systems. \par 
To solve this problem, we exploit the capacity of data-driven approaches \cite{ref_dl_par,Siamese_us_icc,Siamese_us_twc} to obtain an elaborately designed modulation scheme for UWA multicarrier communications, where the discrete representation of baseband waveform is adopted with the aid of Nyquist pulse-shaping. Under the assumption of linear minimum mean square error (LMMSE) equalization at the receiver side, the optimization problem is formulated by considering the fairness of mean square error (MSE) for each data symbol since the total MSE is consistent. To avoid online training and promote generalization ability, the Siamese architecture is involved in forcing consistent results across various channels to guarantee inherent adaptability, which is appropriate for deployment under UWA channels with extremely long feedback delays. Finally, simulation results demonstrate the excellent reliability and robustness of the proposed unified waveform scheme under different UWA channel scenarios by illustrating the bit error rate (BER) performance.\par 
\textit{Notations}: $\mathbf{A}$ is a matrix, $\mathbf{a}$ is a column vector, $a$ is a scalar, $\mathcal{A}$ is a set. $\mathbf{A}^{H}$ and $\mathbf{A}^{-1}$ denote the conjugate transposition and the inverse of $\mathbf{A}$, respectively. $\mathbf{a}(n)$, $\mathbf{a}_{n}$ and $\mathbf{A}_{m,n}$ represent the $n$-th element of the column vector $\mathbf{a}$, the $n$-th column of $\mathbf{A}$ and the $(m,n)$-th entry of $\mathbf{A}$, respectively. $||\mathbf{a}||_{2}$ stands for the $l_{2}$-norm of $\mathbf{a}$ while $||\mathbf{A}||_{F}$ denotes Frobenius norm of $\mathbf{A}$. Finally, the $N$-dimensional identity matrix is denoted as $\mathbf{I}_{N}$. \par
\section{System Model}
Considering the feasibility of digital implementation, the linear combination of Nyquist filters is adopted as the subcarriers, i.e., the system is implemented based on the precoded single-carrier strategy \cite{ref_VBMC}. To be more specific, let $N$ and $\mathbf{F}\in\mathbb{C}^{N\times N}$ respectively denote the number of subcarriers and precoding matrix at the transmitter, which is also known as the digital modulation matrix. The $k$-th subcarrier $g_{k}(t)$ can then be derived as 
\begin{equation}
	g_{k}(t)=\sum_{n=0}^{N-1}\mathbf{F}_{n,k}g(t-nT),
	\label{subcarrier}
\end{equation}
where $T$ and $g(t)$ is the sampling period and the Nyquist pulse of the symbol interval $T$. Let $\{x_{k}|k=0,\cdots,N-1\}$ stand for the data symbols loaded by the corresponding subcarriers, the baseband waveform at the transmitter can then be obtained by the conventional multicarrier scheme as
\begin{equation}
	s(t)=\sum_{k=0}^{N-1}x_{k}g_{k}(t).
	\label{multicarrier}
\end{equation}
For ease of illustration and formulation, let $\mathbf{x}\in\mathbb{C}^{N\times1}$ with $\mathbf{x}(k)=x_{k}$ and substitute \eqref{subcarrier} in \eqref{multicarrier}, $s(t)$ can be rewritten as
\begin{equation}
	s(t)=\sum_{n=0}^{N-1}s_{n}g(t-nT),
	\label{SC}
\end{equation}
where $s_{n}$ denotes the $n$-th time domain sample. Therein, we have $s_{n}=\sum_{k=0}^{N-1}\mathbf{F}_{n,k}x_{k}$, i.e., $\mathbf{s}=\mathbf{Fx}\in\mathbb{C}^{N\times1}$ with $\mathbf{s}(n)=s_{n}$. Without loss of generality, we assume $\mathbb{E}(\mathbf{x})=\mathbf{0}$ and $\mathbb{E}(\mathbf{x}\mathbf{x}^{H})=\mathbf{I}_{N}$ to normalize the transmit power. To guarantee the orthogonality among subcarriers, $\mathbf{F}$ is required to be a unitary matrix, i.e., $\mathbf{F}^{H}\mathbf{F}=\mathbf{I}_{N}$. Then ZP is added with the length of $T_{g}=N_{g}T$ to confront the severe multipath delay spread in UWA scenarios.\par 
After the up-conversion, the continuous passband signal $\tilde{s}(t)=\Re\left\{s(t)e^{j2\pi f_{c}t}\right\}$ is sent from the transmitter to the receiver via the multipath UWA channels, which are known as the wideband time-variant channels with delay-scale spread \cite{ref_TSP_CE_classic,ref_VBMC}. The received passband waveform after appropriate filtering can be derived as \cite{ref_VBMC,ref_TSP_CE_classic,ref_ODSSM}
\begin{equation}
	\tilde{r}(t)=\sum_{p=1}^{P}{A}_{p}\tilde{s}\bigg(t-\big(\tau_{p}-a_{p}t\big)\bigg)+\tilde{w}(t),
\end{equation}
where $P$, $A_{p}$, $a_{p}$, and $\tau_{p}$ represent the number of paths, path amplitude, Doppler scaling factor, and propagation delay associated with the $p$-th path, respectively. The Doppler scaling factor can be further represented as $a_{p}=\frac{v_{p}}{c}$, where $v_{p}$ denotes the velocity with which the $p$-th path length is decreasing while $c$ represents the speed of sound in underwater environments. The maximum delay spread and Doppler scaling spread are denoted as $\tau_{p}\leq\tau_{\max}$ and $|a_{p}|\leq a_{\max}$, respectively. $\tilde{w}(t)$ stands for the additive noise at the receiver. After down-conversion, the baseband received signal can be derived as
\begin{equation}
	r(t)=\sum_{p=1}^{P}h_{p}e^{j2{\pi}a_{p}f_{c}t}s\bigg(\left(1+a_{p}\right)t-\tau_{p}\bigg)+w(t),
	\label{receive_baseband}
\end{equation}
where $h_{p}=A_{p}e^{-j2{\pi}f_{c}\tau_{p}}$ is the equivalent complex gain associated with the $p$-th path while $w(t)$ represents the baseband additive white Gaussian noise (AWGN). $r(t)$ is then sampled at the period $T$ to obtain time domain samples as
\begin{equation}
	\begin{aligned}
		r_{n}=r(nT)=\sum_{n^{\prime}=0}^{N-1}H_{n,n^{\prime}}s_{n^{\prime}}+w(nT),
		\label{element_samples}
	\end{aligned}
\end{equation}
for $n=0,1,\cdots,N+N_{g}-1$, where we have
\begin{equation}
	H_{n,n^{\prime}}=\sum_{p=1}^{P}h_{p}e^{j2{\pi}a_{p}f_{c}nT}g\big(\left(1+a_{p}\right)nT-\tau_{p}-n^{\prime}T\big),
	\label{channelMatrix}
\end{equation}
by combining \eqref{receive_baseband} and \eqref{SC}. \eqref{element_samples} can also be rewritten in vectorized form as $\mathbf{r}=\mathbf{Hs}+\mathbf{w}$ with $\mathbf{r}(n)=r_{n}$, $\mathbf{H}_{n,n^{\prime}}=H_{n,n^{\prime}}$, $\mathbf{w}(n)=w(nT)$, respectively. Let $\sigma^{2}$ represent the variance of AWGN samples, which leads to $\mathbf{w}\sim\mathcal{CN}(\mathbf{0},\sigma^{2}\mathbf{I}_{N+N_{g}})$. For ease of illustration, the notation of $\mathbf{H}_{e}=\mathbf{HF}$ is adopted to denote the equivalent channel matrix between time domain received samples $\mathbf{r}$ and data symbols $\mathbf{x}$, i.e., $\mathbf{r}=\mathbf{H}_{e}\mathbf{x}+\mathbf{w}$, which is employed for the modulation optimization in the following section.\par 
\begin{figure*}
	\centering{\includegraphics[width=0.95\linewidth]{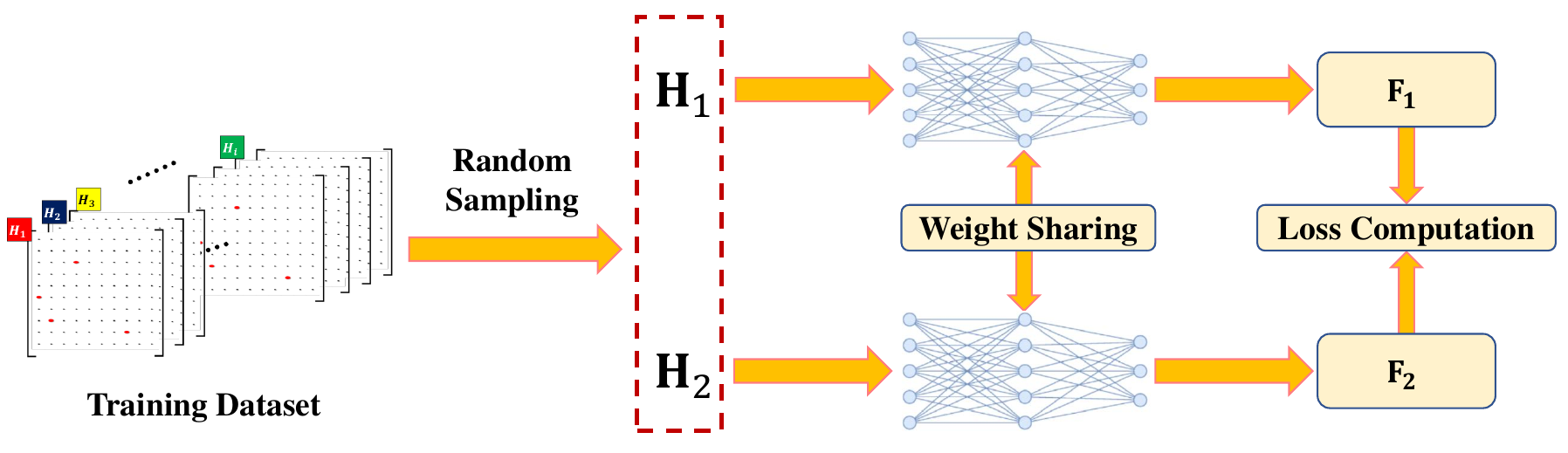}}
	\caption{Schematic of the proposed Siamese-based optimization framework.}
	\label{Fig_Siamese}	
\end{figure*} 
\section{Proposed Scheme}
Theoretical analysis has demonstrated that interference-free transmission is unachievable due to the doubly-dispersive nature of the UWA channels \cite{inter_must}. In such scenarios, the error performance is jointly determined by both the digital modulation $\mathbf{F}$ and the equalization scheme at the receiver. In this paper, we assume that the LMMSE detector is adopted at the receiver considering the complexity and analysis superiority. The fairness of MSE for each data symbol is first investigated to establish the optimization framework. The Siamese neural network is then employed to obtain the modulation design by exploiting the potential of deep learning, whose network design is also introduced in detail. \par 
\subsection{High-Level Idea of the Proposed Scheme}
The modulation scheme plays a significant role in the system performance by reshaping the structure of the equivalent channel $\mathbf{H}_{e}$, whose impact is also determined by the specific receiver designs. In this paper, we adopt the LMMSE equalization and investigate the MSE criteria to optimize the modulation matrix $\mathbf{F}$, which can also be utilized for linear zero forcing (LZF)-based systems. At the receiver, the data symbols are recovered by the LMMSE equalization as
\begin{equation}
	\hat{\mathbf{x}}=(\mathbf{H}_{e}^{H}\mathbf{H}_{e}+\sigma^{2}\mathbf{I}_{N})^{-1}\mathbf{H}_{e}^{H}\mathbf{r}.
\end{equation}
The correlation matrix of the estimation error can then be derived as 
\begin{equation}
	\label{covariance_est}
	\begin{aligned}
		\mathbf{C}_{\mathbf{H}}&=\mathbb{E}\left((\hat{\mathbf{x}}-\mathbf{x})(\hat{\mathbf{x}}-\mathbf{x})^{H}\right)\\
		&=\sigma^{2}\left(\mathbf{H}_{e}^{H}\mathbf{H}_{e}+\sigma^{2}\mathbf{I}_{N}\right)^{-1}\\
		&=\sigma^{2}\left((\mathbf{HF})^{H}\mathbf{HF}+\sigma^{2}\mathbf{I}_{N}\right)^{-1}\\
		&\overset{(a)}{=}\sigma^{2}\left(\mathbf{F}^{H}(\mathbf{H}^{H}\mathbf{H}+\sigma^{2}\mathbf{I}_{N})\mathbf{F}\right)^{-1}\\
		&=\sigma^{2}\mathbf{F}^{H}\left(\mathbf{H}^{H}\mathbf{H}+\sigma^{2}\mathbf{I}_{N}\right)^{-1}\mathbf{F},
	\end{aligned}
\end{equation}
where $(a)$ is derived by employing the unitary property of $\mathbf{F}$. Therefore, we can easily deduce that the total MSE of the LMMSE estimation is independent of the specific modulation scheme because it can be represented as the trace of the correlation matrix, which leads to
\begin{equation}
	\begin{aligned}
		\text{trace}(\mathbf{C}_{\mathbf{H}})&=\text{trace}\Big(\sigma^{2}\mathbf{F}^{H}\left(\mathbf{H}^{H}\mathbf{H}+\sigma^{2}\mathbf{I}_{N}\right)^{-1}\mathbf{F}\Big)\\
		&=\text{trace}\Big(\sigma^{2}\left(\mathbf{H}^{H}\mathbf{H}+\sigma^{2}\mathbf{I}_{N}\right)^{-1}\mathbf{F}\mathbf{F}^{H}\Big)\\
		&=\text{trace}\Big(\sigma^{2}\left(\mathbf{H}^{H}\mathbf{H}+\sigma^{2}\mathbf{I}_{N}\right)^{-1}\Big).
	\end{aligned}
\end{equation}\par
However, it does not mean that the modulation matrix is not required to be optimized from the view of MSE criteria since MSE associated with each data symbol can still be different. Let $\mathbf{e}_{\mathbf{H}}$ denote the diagonal elements of $\mathbf{C}_{\mathbf{H}}$ in \eqref{covariance_est}, i.e., MSE for each data symbol. Adjusting $\mathbf{F}$ can reshape the distribution of $\mathbf{e}_{\mathbf{H}}$,  which can affect the error performance. Since the data symbol with the worst MSE has the major contributions to the overall error rate, we aim to minimize the maximum value of MSE in $\mathbf{e}_{\mathbf{H}}$. On the other hand, the summation of elements in $\mathbf{e}_{\mathbf{H}}$ is constant in each channel realization, which means this principle can be translated into the fairness among MSE for different data symbols, i.e., the optimal distribution of $\mathbf{e}_{\mathbf{H}}$ can be derived as
\begin{equation}
	\label{optimal_MSE}
	\mathbf{e}_{\mathbf{H},\text{opt}}=\frac{1}{\text{trace}(\mathbf{C}_{\mathbf{H}})}\mathbf{1}_{N}.
\end{equation}\par 
At the meantime, the large transmission delay in UWA communications brought by the small velocity of sound (around 1500 m/s) leads to the failure of CSI-aware adjustment of modulation and demodulation schemes, which requires excellent generalization for the designed waveform. As a result, the optimization problem can be formulated by
\begin{equation}
	\begin{aligned}
		&\min_{\mathbf{F}}  && \mathbb{E}_{\mathbf{H}}\Big(f(\mathbf{e}_{\mathbf{H}},\mathbf{e}_{\mathbf{H},\text{opt}})\Big)\\
		&\quad\text{s.t.} 
		&&\mathbf{H}\in \mathcal{H},\\
		&&&\mathbf{F}^{H}\mathbf{F}=\mathbf{I}_{N},\\
	\end{aligned}
	\label{opt_pro}
\end{equation}
where $f(\mathbf{e}_{\mathbf{H}},\mathbf{e}_{\mathbf{H},\text{opt}})$ is the objective function for MSE fairness, which decreases with the norm between $\mathbf{e}_{\mathbf{H}}$ and $\mathbf{e}_{\mathbf{H},\text{opt}}$. $\mathcal{H}$ represents the set of potential channel matrices, whose size determines the ability of generalization for the proposed modulation. Directly solving \eqref{opt_pro} is nearly impossible considering the doubly-dispersive property of UWA channels. Therefore, the potential of deep learning can be exploited to obtain a near-optimal solution. However, \eqref{opt_pro} requires the constant output for all possible channel matrices in $\mathcal{H}$, which indicates that directly deploying a neural network is not enough since it provides different modulation matrices for different channel matrices. To address this issue, the Siamese neural networks are adopted, which is illustrated in detail in the following subsection.\par 
\begin{algorithm}[t]
	\renewcommand{\algorithmicrequire}{\textbf{Input:}}
	\renewcommand{\algorithmicensure}{\textbf{Output:}}
	\caption{The proposed waveform optimization scheme}
	\label{alg_proposed}
	\begin{algorithmic}[1]
		\REQUIRE
		Channel parameters: $\tau_{\max}$, $a_{\max}$, $T$, $N$ and $N_{g}$\\
		Hyper-parameters: $M_{\text{train}}$, $M_{\text{test}}$, $E$, $B$, and $\lambda$.
		\ENSURE
		Discrete representation of the modulation $\mathbf{F}$
		\STATE
		\textit{\% Generate training and test datasets}
		\STATE
		Initialize $\mathcal{H}_{\text{train}}$ and $\mathcal{H}_{\text{test}}$ as the empty set
		\FOR{$m=1:M_{\text{train}}$}
		\STATE
		Generate $\{a_{p},\tau_{p},A_{p}\}_{p=1}^{P}$ randomly
		\STATE
		Compute the channel matrix $\mathbf{H}$ according to \eqref{channelMatrix}
		\STATE
		Add $\mathbf{H}$ to $\mathcal{H}_{\text{train}}$
		\ENDFOR
		\FOR{$m=1:M_{\text{test}}$}
		\STATE
		Generate $\{a_{p},\tau_{p},A_{p}\}_{p=1}^{P}$ randomly
		\STATE
		Compute the channel matrix $\mathbf{H}$ according to \eqref{channelMatrix}
		\STATE
		Add $\mathbf{H}$ to $\mathcal{H}_{\text{test}}$
		\ENDFOR
		\STATE
		\textit{\% Training process}
		\STATE
		Initialize weights of the neural network $\Theta$
		\FOR{epoch$=1:E$}
		\FOR{batch$=1:B$}
		\FOR{sampling iteration=$1:N_{\text{sample}}$}
		\STATE
		Sample $\mathbf{H}_{1}$ and $\mathbf{H}_{2}$ from $\mathcal{H}_{\text{train}}$
		\STATE
		Obtain $\mathbf{F}_{1}$ by processing $\mathbf{H}_{1}$ with weights $\Theta$
		\STATE
		Obtain $\mathbf{F}_{2}$ by processing $\mathbf{H}_{2}$ with weights $\Theta$
		\STATE
		Compute the loss function according to \eqref{loss_sum}-\eqref{MSE_specific}
		\STATE
		Compute the gradient and store
		\ENDFOR
		\STATE
		Update $\Theta$ utilizing the pre-defined optimizer and the reserved gradients
		\ENDFOR
		\IF{stopping criteria}
		\STATE
		Finish the training process
		\ENDIF
		\ENDFOR
		\STATE
		\textit{\% Generate the proposed modulation matrix}
		\FOR{$m=1:M_{\text{test}}$}
		\STATE
		Obtain $\mathbf{H}_{m}\in\mathcal{H}_{\text{test}}$
		\STATE
		Obtain $\mathbf{F}_{m}$ by processing $\mathbf{H}_{m}$ with weights $\Theta$
		\ENDFOR
		\STATE
		$\tilde{\mathbf{F}}=\frac{1}{M_{\text{test}}}\sum_{m=1}^{M_{\text{test}}}\mathbf{F}_{m}$
		\STATE
		Obtain $\mathbf{F}$ from the QR decomposition of $\tilde{\mathbf{F}}$
	\end{algorithmic}		
\end{algorithm}
\subsection{Learning-Enabled Optimization Framework with Siamese Architecture}
The generalization can be translated into the consistent design across various channel matrices in \eqref{opt_pro}. Therefore, the Siamese architecture \cite{Siamese_us_twc} can be utilized to attain a satisfying output, whose structure is illustrated in Fig. \ref{Fig_Siamese}. Let $\mathcal{H}_{\text{train}}$ denote the training dataset, which usually represents the potential channel matrices with pre-defined delay and Doppler scaling spread. Let $\mathbf{H}_{1}$ and $\mathbf{H}_{2}$ represent two distinct channel matrices sampled randomly from $\mathcal{H}_{\text{train}}$. After processing the high-dimensional features through the same neural network whose weights are denoted as $\Theta$, the output modulation design can be obtained as $\mathbf{F}_{1}$ and $\mathbf{F}_{2}$, respectively. The QR decomposition is integrted into the network to force the orthogonality constraints in \eqref{opt_pro}. Finally, the loss function is computed by combining those two parts. On one hand, the modulation should be designed to minimize the objective function in \eqref{opt_pro}. On the other hand, the Euclidean distance between $\mathbf{F}_{1}$ and $\mathbf{F}_{2}$ should be small enough to enable a consistent solution to various channel scenarios. As a result, the loss function is derived by
\begin{equation}
	\begin{aligned}
		\text{Loss}&=\frac{\lambda}{2}\left(f(\mathbf{e}_{\mathbf{H}_{1}},\mathbf{e}_{\mathbf{H}_{1},\text{opt}})+f(\mathbf{e}_{\mathbf{H}_{2}},\mathbf{e}_{\mathbf{H}_{2},\text{opt}})\right)\\
		&+(1-\lambda)q\left(\mathbf{F}_{1},\mathbf{F}_{2}\right),
	\end{aligned}
\label{loss_sum}
\end{equation} 
where $q(\cdot,\cdot)$ is a measurement of distance between two input variables. In this paper, the exact formulation of $f(\cdot,\cdot)$ and $q(\cdot,\cdot)$ is derived according to the corresponding normalized MSE (NMSE) as
\begin{equation}
	\begin{aligned}
		f(\mathbf{e}_{\mathbf{H}},\mathbf{e}_{\mathbf{H},\text{opt}})=\frac{||\mathbf{e}_{\mathbf{H}}-\mathbf{e}_{\mathbf{H},\text{opt}}||_{2}^{2}}{||\mathbf{e}_{\mathbf{H},\text{opt}}||_{2}^{2}},
	\end{aligned}
\label{objective_specific}
\end{equation} 
and
\begin{equation}
	\begin{aligned}
		q\left(\mathbf{F}_{1},\mathbf{F}_{2}\right)=\frac{||\mathbf{F}_{1}-\mathbf{F}_{2}||_{F}^{2}}{N},
	\end{aligned}
\label{MSE_specific}
\end{equation} 
respectively. \par 
As a result, the proposed scheme of waveform optimization can be summarized in \textbf{Algorithm \ref{alg_proposed}}. According to the maximum delay-scale spread, the training and test datasets are first constructed as $\mathcal{H}_{\text{train}}$ and $\mathcal{H}_{\text{test}}$ with the size of $M_{\text{train}}$ and $M_{\text{test}}$ elements, respectively. After that, the weights of the neural network are initialized. The training process contains $E$ epochs and $B$ batches, where $N_{\text{sample}}=\frac{M_{\text{train}}}{2B}$ sampling iteration is included in each batch. During each sampling iteration, we select channel samples $\mathbf{H}_{1}$ and $\mathbf{H}_{2}$ from $\mathcal{H}_{\text{train}}$ and determines the output of network with same weights as $\tilde{\mathbf{F}}_{1}$ and $\tilde{\mathbf{F}}_{2}$, which leads to the corresponding modulation matrix as $\mathbf{F}_{1}$ and $\mathbf{F}_{2}$ by employing the QR decomposition. After that, the loss function is computed according to \eqref{loss_sum}, and the corresponding gradient is stored. At the end of each batch, the weights of the network are updated according to the pre-defined strategies and the reserved gradients. The training process is ceased when the stopping criteria are satisfied, e.g., the loss function on the test dataset has converged or the maximum number of epochs $E$ is reached. Finally, the proposed discrete representation of waveform is generated based on the average output on the test dataset. It is also worth pointing out that thanks to the constraints of \eqref{MSE_specific}, the difference between $\tilde{\mathbf{F}}$ and $\mathbf{F}_{m}$ can be neglected. On the other hand, the QR decomposition can also merged into the network to force the output to satisfy the constraints in \eqref{opt_pro}. It leads to the consistency between the final waveform $\mathbf{F}$ and the output in each channel sample, which ensures the correctness of the proposed learning-inspired scheme.\par 
\subsection{Network Design}
In this subsection, the network structure is briefly introduced, which completes the overall illustration of the proposed scheme. To effectively leverage channel information for generating modulation schemes suitable for underwater communications, we propose a neural network architecture based on QR decomposition. Specifically, the channel matrix is treated as a dual-channel $M \times N $ image input, where the two channels represent the real and imaginary components of the channel matrix, respectively. To thoroughly extract structural features of the UWA channels with both delay and Doppler scaling spread, the flattened channel matrix undergoes sequential processing through three fully connected (FC) layers with Gaussian Error Linear Unit (GELU) activations. The GELU activation further stabilizes gradient propagation compared to conventional rectified linear units (ReLUs). These 3 layers progressively transform the dimensions to $8N$, $4N$, and $4N$, respectively. Finally, the output layer maps the features into a $2N^2$-dimensional vector, which is reshaped into a $ 2\times N \times N$ modulation matrix. Although the output inherently integrates channel characteristics, it lacks explicit system-level design guidance. To address this, the output is reshaped into a $N\times N$ complex matrix. Then the QR decomposition is applied to the synthesized matrix to obtain the unitary modulation matrix as $\mathbf{F}$, which completes the overall network design.\par   
\subsection{Deployment Issues}
During the off-line training, the dataset containing potential channels can be collected by simulation, ray-tracing, experiments, and many other approaches. Since the output of the off-line training enjoys the inherent consistency across various channel conditions thanks to the Siamese structure and appropriate datasets, there is no need to execute online adjustment during the realistic communications, i.e., neither neural networks nor QR decomposition modules are required to be deployed in realistic transceivers. The only difference between the proposed scheme and traditional schemes is the specific waveform, which can be fixed before the transmission. Therefore, the proposed approach still follows the signal processing of conventional digital modulation systems, which adds no extra complexity in practical implementation since the off-line deployment is required only once. To guarantee the stability of the designed waveform scheme, the generation of a suitable training dataset is well worth investigating in future works.\par
\section{Simulation Results}
\begin{table}
	\caption{Relevant Parameters for Dataset Generation}
	\centering
	\label{simulation_para_table}
	\renewcommand\arraystretch{1.3}
	\begin{tabular}{p{15em}p{10em}}
		\hline
		Parameter &
		Typical value\\
		\hline
		Carrier frequency ($f_{c}$)& $12.5$ kHz\\
		Bandwidth & $5$ kHz\\
		Number of subcarriers ($N$)& 256\\
		Length of ZP ($N_{g}$)& 64\\
		Maximum relative speed ($v_{\max}$)& 20 kn\\
		Modulation alphabet & QPSK\\
		Number of paths ($P$)& 8\\
		Size of training dataset ($M_{\text{train}}$)& 20000\\
		Size of test dataset ($M_{\text{test}}$)& 1000\\
		\hline
	\end{tabular}
\end{table}
In this section, the BER performance is evaluated by conducting numerical experiments, whose relevant parameters are presented in Table \ref{simulation_para_table} following the configurations in \cite{ref_DHT_par,ref_dl_par,UWA_OTFS}. A multipath channel model like \cite{ref_TSP_CE_classic,ref_dl_par} is assumed, where the inter-arrival times are distributed exponentially with the mean of $\mathbb{E}\{\tau_{p+1}-\tau_{p}\}=1$ ms. The amplitudes are Rayleigh distributed with the average power decreasing exponentially with delay, where a total decay from the beginning and the end of the ZP is $20$ dB. The raised cosine pulse with a roll-off factor of $0.65$ is employed as $g(t)$ like \cite{ref_DHT_par}. The Doppler scaling factor associated with the $p$-th path is generated by $a_{p}=a_{\max}\cos(\theta_{p})$ independently like \cite{UWA_OTFS_mine}, where $\theta_{p}$ is uniformly distributed over $[-\pi,\pi]$ while $a_{\max}=\frac{v_{\max}}{c}$ with $v_{\max}$ and $c$ denoting the maximum relative speed and the velocity of sound. During the training stage, the maximum number of epochs and the number of sampling iterations are set as $E=2500$ and $N_{\text{sample}}=200$ while the hyper-parameters for the loss computation are set as $\lambda=0.005$. During the training, the Adam optimizer with learning rate of $2\times10^{-4}$ is adopted to implement the weights update for the neural network. After obtaining the modulation matrix, the BER evaluation is conducted to verify the reliability. LMMSE equalization followed by the hard decision is adopted for all modulation schemes including OFDM, where the inter-carrier-interference is fully considered. Finally, all modulation schemes enjoy the same spectral efficiency since the difference lies in the discrete modulation matrix $\mathbf{F}$, e.g., $\mathbf{F}=\mathbf{I}_{N}$ for single-carrier systems.\par 
\begin{figure}
	\centering{\includegraphics[width=0.95\linewidth]{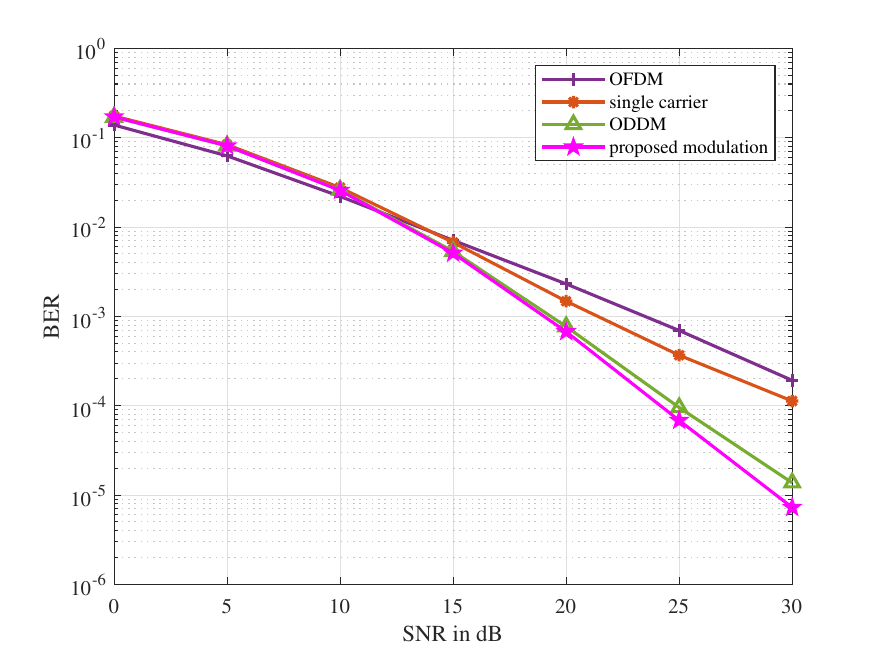}}
	\caption{BER against SNR under training parameters in Table \ref{simulation_para_table}.}
	\label{SimuFig_train}	
\end{figure}  
Fig. \ref{SimuFig_train} illustrates the BER performance against SNR by averaging over 100,000 channel realizations under training parameters in Table \ref{simulation_para_table}. It is obvious that the proposed modulation enjoys the lowest BER across all SNR values. When SNR is 30 dB, BER of the proposed modulation is less than $7\times10^{-6}$ while that of ODDM is still more than $10^{-5}$. The performance degradation for ODDM is mainly due to the mismatch between the narrowband assumptions and wideband effect caused by Dopper scaling factors in UWA scenarios. However, it can still achieve better BER than OFDM and single-carrier systems if full LMMSE is adopted for all schemes.\par 
\begin{figure}
	\centering{\includegraphics[width=0.95\linewidth]{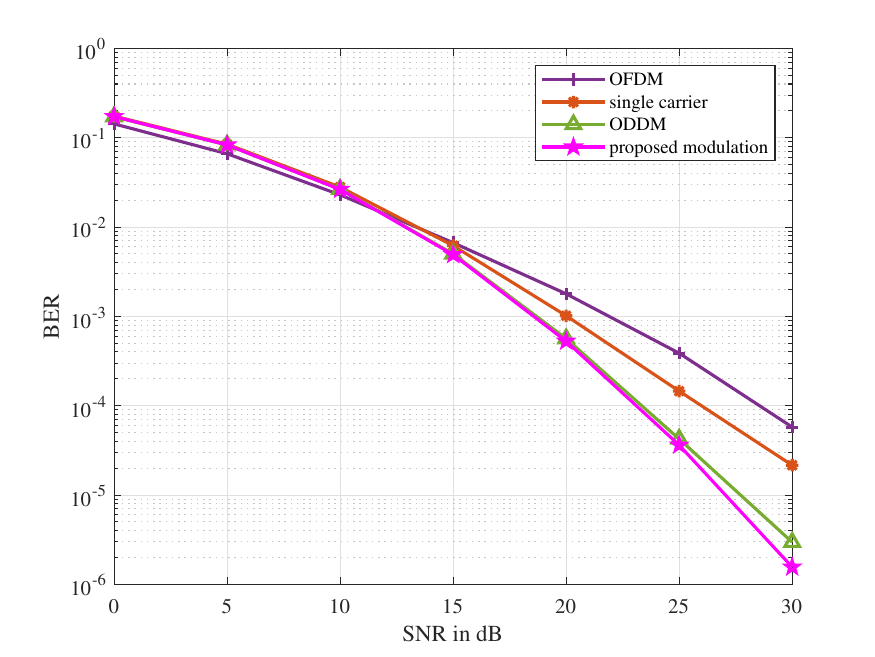}}
	\caption{BER versus SNR under simulation parameters different from Table \ref{simulation_para_table}.}
	\label{SimuFig_robust}	
\end{figure}
The robustness of the proposed modulation is then presented in Fig. \ref{SimuFig_robust} by plotting BER against SNR under parameters different from Table \ref{simulation_para_table}, where we set $f_{c}=35$ kHz, $P=15$, $v_{\max}=10$ kn, respectively. The other parameters are consistent with those in Table \ref{simulation_para_table}. Due to the parameter variation, the performance superiority beyond other modulation schemes degrades compared with the results in Fig. \ref{SimuFig_train}. Nevertheless, the proposed modulation strategy still reveals the best reliability among all schemes. When SNR is 30 dB, the BER of the learned modulation scheme is about $1.8\times10^{-6}$, which is less than 58\% of that in ODDM systems. It is also worth pointing out that by considering the LMMSE equalization with the accurate channel model, ODDM still outperforms conventional OFDM and single carrier systems, which indicates that ODDM can also serve as a qualified candidate for UWA communication even though the Doppler scaling factors should be considered rather than the narrowband Doppler shift.\par 
\section{Conclusion}
In this paper, the modulation was optimized to realize the reliable UWA transmission with severe delay-scale spread, where we adopted the discrete representation by employing the Nyquist pulse-shaping. To enhance the error performance, the optimization objective was first formulated by investigating the fairness among the MSE of each data symbol since the total MSE kept consistent due to the unitary property of orthogonal modulation and LMMSE equalizer at the receiver. The Siamese architecture was then adopted to obtain the unified modulation representation across various UWA channels by exploiting the capacity of deep learning. Both the loss function and the network structure were designed elaborately. Simulation results demonstrated the appreciable performance of the proposed waveform design thanks to MSE fairness and the generalization empowered by the Siamese architecture. For future research directions, it is valuable to consider the relevant channel estimation issues and more realistic channel datasets.\par 
\appendices
\section*{Acknowledgment}
This work was supported by the National Natural Science Foundation of China under Grants 624B2079 and 62271284.
\vfill\pagebreak
\bibliographystyle{IEEEtran}
\bibliography{ref-sum}

\end{document}